\documentclass[fleqn,10pt]{wlscirep}
\title{Activity induced delocalization and freezing in self-propelled systems}

\author[1,*,+]{Lorenzo Caprini}
\author[2,+]{Umberto Marini Bettolo Marconi}
\author[3,+]{Andrea Puglisi}
\affil[1]{Gran Sasso Science Institute (GSSI), Via. F. Crispi 7, 67100 L’Aquila, Italy.}
\affil[2]{Scuola di Scienze e Tecnologie, Università di Camerino - via Madonna delle Carceri, 62032, Camerino, Italy.}
\affil[3]{Istituto dei Sistemi Complessi - CNR and Dipartimento di Fisica, Universit\`a di Roma Sapienza, P.le Aldo Moro 2, 00185, Rome, Italy}

\affil[*]{lorenzo.caprini@gssi.it}

\affil[+]{these authors contributed equally to this work}

\usepackage{graphicx} \usepackage{bm,ulem,color,textcomp}
\usepackage{amsmath,amsfonts,amssymb} 

\usepackage{subfig}
\usepackage{graphicx}

\usepackage[justification=justified, format=plain]{caption}

\usepackage{comment}

\usepackage{soul} 

\begin{abstract}
We study a system of interacting active particles, propelled by
colored noises, characterized by an activity time $\tau$, and confined
by a single-well anharmonic potential.  We assume
  pair-wise repulsive forces among particles, modelling the steric
  interactions among microswimmers. This system has been
  experimentally studied in the case of a dilute suspension of Janus
  particles confined through acoustic traps.  We observe that already
in the dilute regime - when 
  inter-particle interactions are negligible - increasing
the persistent time, $\tau$, pushes the particles away
from the potential minimum, until a saturation distance is reached. We
compute the phase diagram (activity versus interaction length),
showing that the interaction does not suppress this delocalization
phenomenon but induces a liquid- or solid-like structure in the
densest regions.   Interestingly a reentrant behavior is
  observed: a first increase of $\tau$ from small values acts as an
  effective warming, favouring fluidization; at higher values, when
  the delocalization occurs, a further increase of $\tau$ induces
  freezing inside the densest regions.   An approximate analytical
scheme gives fair predictions for the density profiles in the weakly
interacting case. The analysis of non-equilibrium heat fluxes reveals
that in the region of largest particle concentration equilibrium is
restored in several aspects.

\end{abstract}

\begin{document}


\flushbottom
\maketitle
\thispagestyle{empty}

\section*{Introduction}
Recently the theorists' attention has been attracted by the study of
so-called self-propelled particles~\cite{bechinger2016active,ramaswamy2010mechanics,marchetti2013hydrodynamics} in the context of active matter.
Typical experimentally accessible examples come from biological systems: swimming bacteria, such as the E. Coli~\cite{Berg2004Ecoli}, unicellular protozoa~\cite{Blake1974ciliary} and spermatozoa~\cite{Wolley2003Spermatozoa} but also more complex systems such as actin filaments~\cite{bausch2011}, active nematics~\cite{Sague2016}, living tissues~\cite{Poujade2007} or the so-called motor-proteins~\cite{Dogic2012}.
Moreover, artificially realized micro-swimmers, such as self-propelled
Janus particles~\cite{Lattuada2011Synthesis,Walther2013Janus} and
colloidal particles immersed in a bacterial
suspension~\cite{Maggi2014Equipartition}, have been shown to behave as
active systems.  All these examples show common features both at the
level of the single particle trajectory~\cite{FodorMarchetti2018}, and
at the collective level, which cannot be captured by an
  equilibrium Brownian motion model.  Regardless of
  their nature, these systems propel themselves in some space
  direction for a finite time, by employing different
  mechanisms. Typically, biological systems employ mechanical tools,
  such as Cilia or Flagella, or complex chemical reactions. Active
  colloids are typically activated through
  light~\cite{Jiang2010,Buttinotti2012}, which injects energy into the
  system, or chemically through the decomposition of hydrogen
  peroxide~\cite{Paxton2004,Howse2007}.  Independently
  on their origins, on one hand, an isolated
self-propelled particle displays an anomalously long
persistent motion: In the potential-free case, its
  orientation - i.e. the active force - and so its velocity direction
  remains constant for times much longer with respect to those allowed
  by an equivalent thermal system, where activity is replaced by
  ordinary diffusion process with the same amplitude.  Of course, at very
long times - when velocity correlations have decayed - normal
effective diffusion is recovered when active
particles are not confined.  On the other hand, a
suspension of interacting active particles shows interesting
collective phenomena such as the so-called motility induced phase
separation
(MIPS)~\cite{fily2012athermal,buttinoni2013dynamical,bialke2015active,cates2015motility,palacci2014nature,Redner2013,Pohl2014,Gonnella2015motility}
or dynamical ordering phenomena such as
flocking~\cite{Vicsek2012Flocking}. All these phenomena cannot be
explained through an equilibrium approach,  i.e. in terms of a
  Maxwell-Boltzmann distribution.  For this reason, a series of
simplified models have been recently proposed, in particular, the Run
and Tumble model~\cite{Tailleur2009Sedimentation,Nash2010Run,tailleur2008statistical}
and the Active Brownian Particles (ABP)
model~\cite{romanczuk2012active,Hagen2011,Zott2016}: the connection
between these two modelizations was discussed
in~\cite{solon2015active, cates2013active}, showing a good consistency
between them, at least in a range of values of the control parameters.
Since the two-time correlation of stochastic activity force in the
ABP, averaged over the angular degree of freedom, has an exponential
shape, the Active Ornstein-Uhlenbeck Particles (AOUP) model was
introduced, as the simplest model with such time-correlation
behavior~\cite{farage2015effective,szamel2014self,maggi2015multidimensional,Gompper2018,flenner2016nonequilibrium}.
Despite its apparent simplicity, many aspects of the active
phenomenology were
reproduced~\cite{maggi2015multidimensional,fodor2016far}, providing
consistency with this model. The possibility in AOUP of obtaining clear
theoretical results may lead to new predictions which may trigger
future experimental investigations.

With this aim, here we implement numerical
simulations~\cite{ToralColen_book} of interacting AOUP particles
within a confining single-well potential, reproducing a
``delocalization'' phenomenon, i.e. the escape of particles far from the
potential minimum, recently observed in experiments with Janus
particles.  In particular, in~\cite{Takatori2016Nature} the system was
dilute enough to make inter-particle interactions
negligible.  At variance with~\cite{Hennes2014}, our
  model considers a constant mobility and neglects any kind of
  hydrodynamic interactions, supposed to be small. Moreover, we do not
  involve any alignment and consider only pairwise repulsive
  potentials to model the steric repulsion among the spherical
  microswimmers.  Our study first demonstrates that delocalization
increases with activity and is robust also in the presence of
interactions, at least up to a certain effective density. We also
reveal a complex interplay between interactions and activity, inducing
a freezing phenomenon which is consistent with the one observed with
ABP particles in~\cite{Bialke2012Freezing}. The relative simplicity of
the AOUP model allows understanding the physical origin of both
delocalization and freezing.  In particular, an approximation method,
the so-called Unified Colored Noise Approximation (UCNA),
well reproduces the density profiles, offering a simple principle for
determining the density in the case of non-interacting
  particles subject to external fields.  An interesting observation
concerns the role of detailed balance (DB)~\cite{Gardiner, risken}
which is locally satisfied only in regions of space having the highest
probabilities of being occupied, while in the remaining regions DB is
violated and the local velocity distribution displays strongly
non-Gaussian shapes.


\section*{Model and numerical Results}
As mentioned in the Introduction, one of the most popular models describing self-propelled particles is ABP. The microswimmers are approximated as points and the hydrodynamic interactions due to the fluid feedback are neglected. The self-propulsion mechanism 
is represented by a force of amplitude $v_0$ and direction $\hat{\mathbf{e}}_i$. For instance,  in two dimensions $\hat{\mathbf{e}}_i$ is a vector of component $(\cos{\theta_i}, \sin{\theta_i})$, being $\theta_i$ the orientational angle of particle $i$.
Therefore, the radial component of the activity is assumed to be constant. The ABP dynamics describing a suspension of $N$ particles in a two-dimensional system reads:
\begin{equation}
\label{eq:ABPdynamics}
\begin{aligned}
\gamma\dot{\mathbf{ x}}_i &=\mathbf{F}_i + \gamma\sqrt{2D_t}\boldsymbol{\xi}_i + \gamma v_0 \hat{\mathbf{e}}_i \\
\dot{\theta}_i &=\sqrt{2D_r}w_i
\end{aligned}
\end{equation}
where $\boldsymbol{\xi}_i$ and $w_i$ are independent white noises
(i.e. $\delta$-correlated in time and with zero average).  $D_r$ is
the rotational diffusion coefficient, which states the typical time
associated to the activity directional change, $\tau_r \sim 1/D_r$.
$F_i$ is the total force acting on the particle $i$, which can be
decomposed as $\mathbf{F}_i = -\nabla_i U(\mathbf{x}_i ) - \nabla_i
\Phi(\mathbf{x}_1 , ... \mathbf{x}_N )$, i.e. into the force due to
the external and to the interaction pairwise potential,
respectively. We call $l$ and $R$, respectively, the typical length of
$U$ and $\Phi$, such that $\Phi = \sum_{i<j} \phi(|\mathbf{x}_i - \mathbf{x}_j|/R)$ and $U=U(\mathbf{x}/l)$. For the sake of simplicity, $l$ is set to one in the numerical study.
The parameters $\gamma$ and $D_t$ denote the solvent
viscous damping and the bare diffusivity due to thermal fluctuations
(i.e. in the absence of forces and activity).  
Notwithstanding its clarity, deriving
further analytical predictions for the ABP model may be difficult even in
simple cases.
The form of the  autocorrelation function, $\langle \hat{\mathbf{e}}_i(t) \cdot \hat{\mathbf{e}}_j(t') \rangle$, of the orientational d-dimensional vector $\hat {\mathbf{e}}_i$ is well known in the theory of rotational diffusion of polar molecules \cite{bookzwanzig2001}.
For generic $d$, averaging over the angular distributions at time $t$ and $t'$, we simply obtain $\langle \,\langle \hat{\mathbf{e}}_i(t) \cdot \hat{\mathbf{e}}_j(t') \rangle\,\rangle_{\Omega}= \exp{(-|t-t'| D_r (d-1))}\delta_{ij}$, being $\langle \cdot \rangle_{\Omega}$ the average over the angular degrees of freedom.
For this reason, as already mentioned in the
Introduction, the AOUP model has been introduced as a surrogate able
to capture the ABP phenomenology. Indeed, the AOUP model is perhaps
the simplest model which exhibits the same two-time correlations
matrix as the ABP. In the AOUP one replaces $v_0\hat{\mathbf{e}}_i
\rightarrow \mathbf{u}^a_i$ in Eq.\eqref{eq:ABPdynamics}, where each
component of $\mathbf{u}^a_i$ evolves as an independent
Ornstein-Uhlenbeck process.  AOUP dynamics reads:
\begin{equation}
\label{eq:AOUPmodel}
\begin{aligned}
&\gamma\,\dot{{\bf x}}_i= {\bf F}_i({\bf x}_1, ... , {\bf x}_N) + \gamma {\bf u}^a_i + \gamma\sqrt{2 D_t}\boldsymbol{\xi}_i,\\
&\tau\,{\bf \dot{u}}^a_i= -{\bf u}^a_i+\sqrt{2D_a}\boldsymbol{\eta}_i,
\end{aligned}
\end{equation}
where $\boldsymbol{\eta}_i$ is a d-dimensional noise
  vector, whose components are $\delta$-correlated in time and have
  unit variance and zero mean.  In this approximation
the term $\gamma \mathbf{u}^a_i$ represents the self-propulsion
mechanism, the internal degree of freedom which converts energy into
motion, such that $\langle \mathbf{u}^a_i (t)\cdot \mathbf{u}^a_j (t' )\rangle =  D_a/\tau
\exp{(-|t-t'|/\tau)} \delta_{ij} d$. Finally, the non-equilibrium
parameters $\tau$ and $D_a$ are, respectively, the persistence time
and the diffusivity due to the activity, which is usually some order
of magnitude larger than $D_t$, an approximation often
  employed also in the ABP model.  The identification of
  the two correlations matrices imposes relations among the
  coefficients, namely in $v_0^2/d =  D_a/\tau$ and $D_r (d-1) = 1/\tau$.
  Since the third, fourth, and so on, correlation matrices are in
  general non-trivial in the ABP, the AOUP model can be considered as
  its effective Gaussian approximation. Moreover, the unitary
  constraint of activity is removed meaning that the radial component
  of the activity has itself a dynamics. As revealed by extensive
  numerical studies, these approximations seem not to be particularly
  relevant in order to recover the self-propelled particles
  phenomenology and for these reasons one could claim the
  possibility to consider the AOUP as a basic model itself and not
  simply as an ABP approximation. 

We point out that in the potential-free model there are two natural
temperatures: the
solvent temperature $T_b=\gamma D_t$ and the effective active
temperature $T_a=\mu \langle u^2 \rangle = \mu D_a/\tau = \gamma D_a$,
where we have defined the effective mass $\mu=\gamma \tau$ (see
below). We fix the value of $\gamma=1$
and inspired to the connection between the AOUP and the ABP model~\cite{Gompper2018} - we also fix the ratio $D_a/\tau=10$ that is the
variance of the self-propulsion velocity. This protocol allows us to
use a single parameter, $\tau$, to tune the relevance of activity in
the system. In fact, taking the limit $\tau\rightarrow 0$ leads to
$T_a \ll T_b$, providing a vanishing contribution with respect to the
thermal noise.  On the contrary, at large values of $\tau$ one
has $T_a \gg T_b$: self-propulsion becomes important and the
thermal bath can be neglected. We restrict to this second regime, where the
system temperature $T_a$ has a limited significance since it
represents the temperature of the system only in few specific cases
discussed below and in the Supplementary Information (SI). In general, the system is out of equilibrium and many
of its statistical properties are hardly comparable to a thermal
system.

A well-known result for this model concerns the existence of
MIPS in the large activity regime, when $U \equiv 0$ and 
$\Phi$ is  given by the sum of pairwise repulsive potentials~\cite{fodor2016far}. In this work, 
  the particles are confined by an external radial potential, $U(r) \propto
r^{2n}$ with $r=|{\bf x}|$. We choose $n>1$ since the case $n=1$ -
when thermal noise and particle-particle interactions are negligible -
is trivial even at $\tau>0$, corresponding to a Gibbs density
distribution $\sim e^{-U(r)/T_{eff}}$ with some temperature
$T_{eff}$ (see discussion after Eq.~\eqref{eq:xv_model} and SI). If $n>1$ and
$\tau>0,$ DB is broken and the steady phase-space distribution is not amenable to a simple representation in terms of $U(r)$. 
In the presence of an
external potential, a useful dimensionless parameter 
  can be defined:
 $$\nu=\frac{\tau U''(l)}{\gamma}.$$
 It represents the ratio
between the persistence time and the relaxation time due to the
external force:  $\nu$ is a relative measure of the activity in our system and determines how far from equilibrium is the
system. Indeed, when $\nu \lesssim 1$, the
relaxation time of the active force is smaller than the typical time
over which a significant change of the microswimmer position, due to
the potential, occurs:  thus from Eq.\eqref{eq:AOUPmodel}
    we have $\boldsymbol{u}^a_i \approx
    \sqrt{2D_a}\boldsymbol{\eta}$. In this case, one recovers an
    equilibrium-like picture, which can be explained in terms of the
    effective temperature, $T_a=\gamma D_a$ (SI for more
      details). When $\nu \gg 1$, the situation dramatically changes:
    we have to take into account the dynamics of both degrees of
    freedom in Eq.\eqref{eq:AOUPmodel} and we expect significant
    departures from an equilibrium-like picture.   Note that keeping
  fixed the strenght of the activity, $D_a/\tau=v_0^2$,
  $\gamma$ and $U(r)$, one has that both $T_a$ and $\tau$ are
  proportional to $\nu$.

\subsection{Phase diagram: delocalization and induced freezing}
\begin{figure}[!t]
\centering
\includegraphics[width=0.95\linewidth,keepaspectratio]{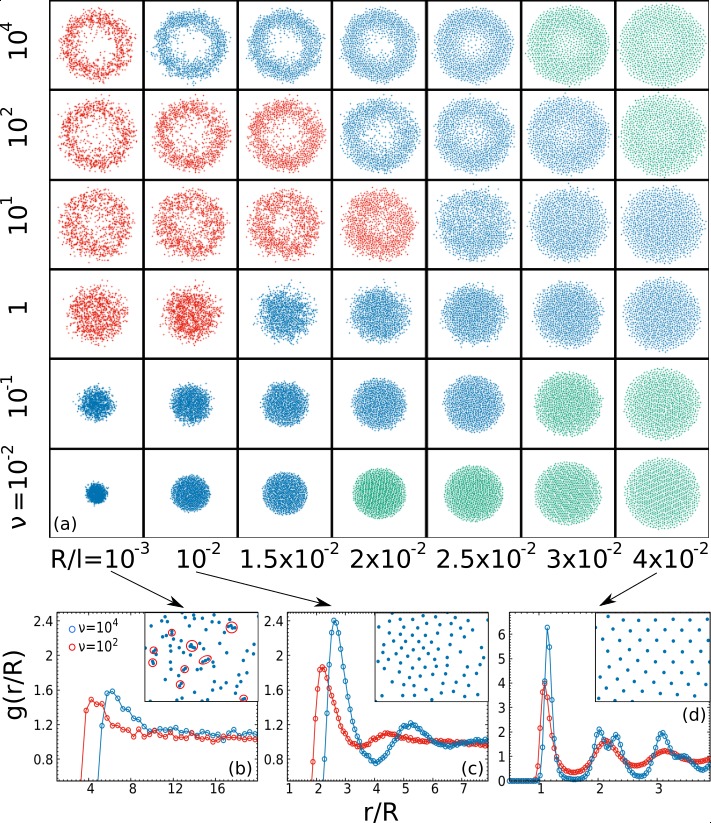}
\caption{Top: Phase diagram illustrated by simulation snapshots as a
  function of $R/l$ and $\tau$. Colors indicate the internal density
  structure (see SI for details): gaseous (red), liquid (blue), solid (green). Bottom:
  $g(r/R)$ for $R=10^{-3}, 10^{-2}, 3\cdot10^{-2}$, and for two
  different values of $\tau=1 , 10^2$, respectively red and blue
  dots. Each box is realized confining $N=10^3$ particles through the
  interaction potential $\Phi\sim  \sum_{i<j} R^4/|{\bf x}_{ij}|^4$. Parameters:
  $n=2$, $D_a/\tau=10^2$ and $D_t=10^{-5}$.}\label{Fig:Snapshot_AOUP}
\end{figure}

In Fig.~\ref{Fig:Snapshot_AOUP} we display pictorially the phase
diagram of a system in 2 dimensions, varying $\nu$ and the rescaled
interaction radius $R/l$,  (keeping fixed the number of
  particles and the external potential), which play the role of
  control parameters. Through $R/l$ we control the excluded volume of
  the system, while through $\nu$ we tune the relevance of the
  activity ingredient. 

Considering the non-interacting regime - $\Phi=0$ or
  equivalently $R/l$ small enough as in the left column of
  Fig.~\ref{Fig:Snapshot_AOUP}~(a) - the equilibrium-like regime for
  $\nu \lesssim 1$ is consistent with a Brownian-like picture and does
  not reveal any surprises: particles accumulate around the minimum of
  the potential, exploring an effective average volume determined just
  by the interplay between the external potential and the random force. Indeed, the system has effective temperature
  $T_a$, and no far-from-equilibrium physics is involved.
In the non-equilibrium regime, namely $\nu \gtrsim 1$ in
  Fig.\ref{Fig:Snapshot_AOUP}~(a), the area close to the potential
  minimum empties and the system shows strong delocalization in favour
  of a peripheric (annular in 2d) region at an average distance $r^*$
  from the origin.  At large values of $\nu$, $r^*$ appears to
  saturate and a further increase of $\nu$ just produces a dynamical
  effect, leading to a slowdown of the particles (see SI for
  details). This phenomenology reproduces the experimental result
  obtained in~\cite{Takatori2016Nature} for Janus particles inside an
  acoustic trap with negligible interactions.

Let us to discuss the interacting case, that is when $R/l$
  is not negligible.  The equilibrium-like regime, when $\nu\lesssim
  1$ in Fig.\ref{Fig:Snapshot_AOUP}~(a), can be again understood in
  terms of a Brownian picture. Indeed, the system has temperature
  $T_a$, regardless of $R/l$, and we recover the three
  equilibrium-like aggregation phases, as expected: a
    dilute-phase (or gas), where interactions between particles are
    rare and the volume explored by the particles is only controlled
    by the random force; a solid-like phase, where the random force is
    very small compared to the inter-particle interactions and
    produce only oscillations around the almost-fixed particles
    positions; and finally; an intermediate liquid-like phase where both these terms
    are relevant and produce a correlated and complex dynamics.  
   These different internal structures can be roughly
    identified by the study of the pair correlation
    function~\cite{lowen1994review}, $g(r)$, which is estimated by taking into account a region
approximately uniform in density, in the densest part of the system
(namely the annular region): in the dilute regime
    $g(r)$ is flat or "quasi"-flat, in the liquid one $g(r)$ displays,
    some peaks before approaching to one and finally in the solid
    regime these peaks become more pronounced, showing the typical
    structure of hexagonal lattice (in 2D with radial inter-particles
    interactions).
In all the equilibrium-like aggregation phases the increasing of $\nu$ produces an expected fluidization phenomenon, which can be easily understood in terms of the effective temperature, $T_a \propto \nu$.
In particular, in the liquid-like regime, as shown in the first  two left columns of Fig.\ref{Fig:Snapshot_AOUP}~(a), the increase of $\nu$ enhances the effective volume occupied - when the excluded volume becomes negligible compared to noise-fluctuation -, leading to the transition from the liquid-like to the gas-like structure. 
In the solid-like regime - last  two  right columns in Fig.\ref{Fig:Snapshot_AOUP}~(a) -, the interactions are very strong and the effectively occupied volume is determined by the balance between the inter-particle repulsion and the confinement due to the external potential. In this case, the increase of $T_a$ leads only to the fluidization of the internal structure of the system, determining the transition from a solid-like to a liquid-like structure.

Restricted to $\nu \gtrsim 1$, the delocalization
phenomenon persists when the interaction radius $R/l$ increases, as
shown in Fig.~\ref{Fig:Snapshot_AOUP}~(a). In that case, it is
interesting to analyze the internal structure of the system,
 exploiting analogies and differences with respect to the
  equilibrium picture.  In this regime of $\nu$, this
  analysis leads to the identification of non-equilibrium aggregation
  phases which resemble the equilibrium scenario but with important
  differences, which already emerges from the static structure.
 Indeed, non-equilibrium effects
manifest themselves in two ways: 1) in the dilute case - i.e. left column of Fig.~\ref{Fig:Snapshot_AOUP}~(a) -
a peak at $r\sim R$ appears in
the $g(r)$, not expected for dilute Brownian particles at the same conditions in terms of density and temperature (Fig.~\ref{Fig:Snapshot_AOUP}~(b) and SI for details): this
is likely to be similar to that observed in~\cite{maggi2015multidimensional}; 2) increasing $R/l$, the system displays liquid-like and solid-like structures but
with evident shifts in position and intensity with respect to an
equilibrium structure with comparable average energy per particle and density, as shown in Fig.~\ref{Fig:Snapshot_AOUP}~(c) and (d) (see
SI for details). At large $R/l$ - but still far from close packing -
the system freezes into an almost periodic lattice structure just by
increasing $\nu \sim \tau$. This analysis suggests that a purely
dynamical quantity, the persistence time, {\color{red}$\tau$}, can produce a dramatic
change in the internal structure of the system.  Finally, when the
interaction radius $R/l$ brings the system to an effective close
packing, the radial delocalization is completely suppressed and the system
comes back to a homogeneous phase with ordered (solid-like) internal
structure.  In this regime, inter-particle interactions dominate compared to active forces, which are completely negligible.

 Summarizing, for all the explored values of $R/l$, namely in all the aggregation phases, our numerical study suggests a reentrant behavior of the structural properties of the system induced by $\nu$. The first fluidization, explained by the effective temperature approach, is followed by an induced far-from-equilibrium freezing for $\nu\gtrsim1$, which requires a more subtle analysis to be understood.
The discussion, at least regarding the delocalization phenomenon, remains qualitatively valid in three dimensions.

\section*{Theoretical approach}
In order to make analitycal progress, it is common to map Eq.~\eqref{eq:AOUPmodel} onto a different system, going from the
description in the variables $({\bf x}_i, {\bf u}^a_i)$ to $({\bf x}_i, {\bf v}_i \equiv \dot{\bf x}_i)$,  i.e. considering the evolution of the coarse-grained velocity of each particle instead of their the activity.
When the thermal noise is negligible ( i.e. $D_t \ll D_a$), deriving with respect to time Eq.\eqref{eq:AOUPmodel} and eliminating ${\bf u}^a_i$ in favor of ${\bf v}_i$, leads to\cite{marconi2015towards} (see also SI):
\begin{subequations}
\label{eq:xv_model}
\begin{align}
\dot{{x}}_{i\alpha}&={v}_{i\alpha}\\
\mu \,\dot{{v}}_{i\alpha} &= -\gamma{\Gamma}^{\alpha \gamma}_{ik} { v}_{k\gamma} + { F}_{i\alpha} + \gamma\sqrt{2D_a}\eta_{i\alpha}, \label{eq:v}\\
\Gamma^{\alpha\gamma}_{ik} &=  \left(\delta_{ik}+\frac{\tau}{\gamma} \frac{\partial}{\partial x_{i\alpha}} \frac{\partial}{\partial x_{k\gamma}} (\Phi + U)\right). 
\end{align}
\end{subequations}
where we use Latin and Greek indices for indicating the $N$ particle
and for the $d$ components of the particle coordinates, respectively.
 We point out that this mathematical passage can be
  considered only as a change of variables and thus does not involve
  any approximations.   Moreover, $v_{i\alpha}$ is not
  the real velocity of such a particle but has to be interpreted just
  as a coarse-grained velocity, i.e. $\dot x_{i\alpha}$ by definition,
  where $x_{i\alpha}$ is the position of the overdamped dynamics,
  i.e. such that timescales of molecular interaction and inertia relaxation are filtered
  out.   The original
  over-damped dynamics of each particle is mapped onto the
  under-damped dynamics of a particle immersed into a fictitious bath
  with its effective diffusion coefficient, related to the activity
  parameters.  The non-equilibrium feature of such a dynamics is fully
  contained in the space-dependent, $(d\cdot N)^2$ dimensional
  friction matrix, $\Gamma$, which naturally produces a violation of
  the Fluctuation Dissipation Relation.    The dynamics of one
particle is coupled to all the degrees of freedom through both the
interaction potential and $\Gamma$.  When particle-particle
interactions are negligible, $\Gamma$ reduces to a $d$-block diagonal
matrix, which provides just a coupling among the different components
of the dynamics of a single particle.  In this case, when $\nu \ll 1$,
the $\Gamma$ matrix reduces to a spatially homogeneous matrix and the
system reaches a Gibbs steady state $\sim \exp(-H/T_{eff})$ with
$H=\mu |{\bf v}|^2/2 + U({\bf x})$ and $T_{eff}=T_a$, meaning that
$T_a$ can be identified as the effective temperature of the system
\cite{Puglisi2017Temperatures,Cugliandolo2011EffectiveT}. The
peculiarity of the case $n=1$ emerges in the dilute
  regime since $\Gamma$ is constant for all $\nu$ and
$T_{eff}=T_a(1+\nu)^{-1}$.  In the case $n>1$ and non-negligible
$\nu$, only approximations of the stationary
pdf~\cite{fodor2016far,marconi2017heat,marconi2015velocity} are known.

The representation of the dynamics given by the
Eqs.\eqref{eq:xv_model} sheds some light on both freezing and
delocalization phenomena observed above.  The freezing can be
understood by the slowing down induced by the increase of $\Gamma$,
determined by the internal forces among active particles in the large
persistence regime.  The radial delocalization phenomenon which is
observed even in the presence of negligible interactions can be
physically understood as follows: the effective damping coefficient,
$\Gamma(x)/\tau$, is small near the minimum of the potential well and
increases as $x$ departs from it. Therefore, particles with $x \approx
0$ can attain large velocities and leave the region, while for $x$
large enough they reduce their ``effective speed'', $v$, for the
combined effects of viscous damping and the external force. 

\subsection*{UCNA approximation}
To make this argument quantitative we employ the unified colored noise
approximation (UCNA).
 UCNA was developed first time in~\cite{hanggi1995colored,h1989colored} in the context of electric fields with a correlated noise, but the methodology has been adapted to interacting active particles systems in~\cite{marconi2015towards}.
 This approximation consists in an effective equilibrium approach which predicts the spatial distribution of the particles in terms of an effective potential, which involves the derivatives of $U+\Phi$. The prediction can be derived by dropping the inertial term in Eq.\eqref{eq:xv_model} in the limit of vanishing current, or by performing the usual adiabatic elimination in the FP-equation \cite{Risken?}. Its derivation is reviewed in the SI and the final result reads:
\begin{equation}
\begin{aligned}
&p_{u}({\bf x}_1, ... , {\bf x}_N) \propto e^{-H_{u}({\bf x}_1, ... , {\bf x}_N)/T_a}, \\
&H_{u}=\Phi+U+\frac{\tau}{2\gamma}\sum_{i\alpha}\left(F_{i\alpha}\right)^2-D_a\gamma\log{|\det \Gamma_{ik}^{\alpha \gamma}|}. 
\end{aligned}
\end{equation}
 In spite of the fact that the UCNA is derived under the assumption of vanishing currents and thus restores the DB, at least in some regimes
it is able to capture many interesting aspects of the observed phenomenology of self-propelled particles.

In order to assess this approximation, we consider a one-dimensional
system of non-interacting particles and show, in Fig.~\ref{fig:EnergyUCNA}~(a), the comparison between
the numerical probability density in space, $p(x)$, and $p_u(x)$.
Remarkably, the effective potential $H_{u}$ takes the shape of a
double well which fairly reproduces the numerical simulations. The
comparison is optimal when $\tau\ll 1$, and gives fair quantitative
information for the location of the density maxima also when
$\tau\gg 1$. In particular, $p_u$ correctly predicts the accumulation
in some regions, depending on $\tau$, but it undererrates the
probability of finding a particle in the bottom of the well, for large
$\tau$. This double-well effective phenomenology may be also related to the results
obtained in~\cite{caprini2018LinearResponse}, explaining why the time-dependent response
function of the system shows two different characteristic times for large values
of the activity.

\subsection*{Hydrodynamics}
\begin{figure}[!t]
\centering
\includegraphics[width=1\linewidth,keepaspectratio]{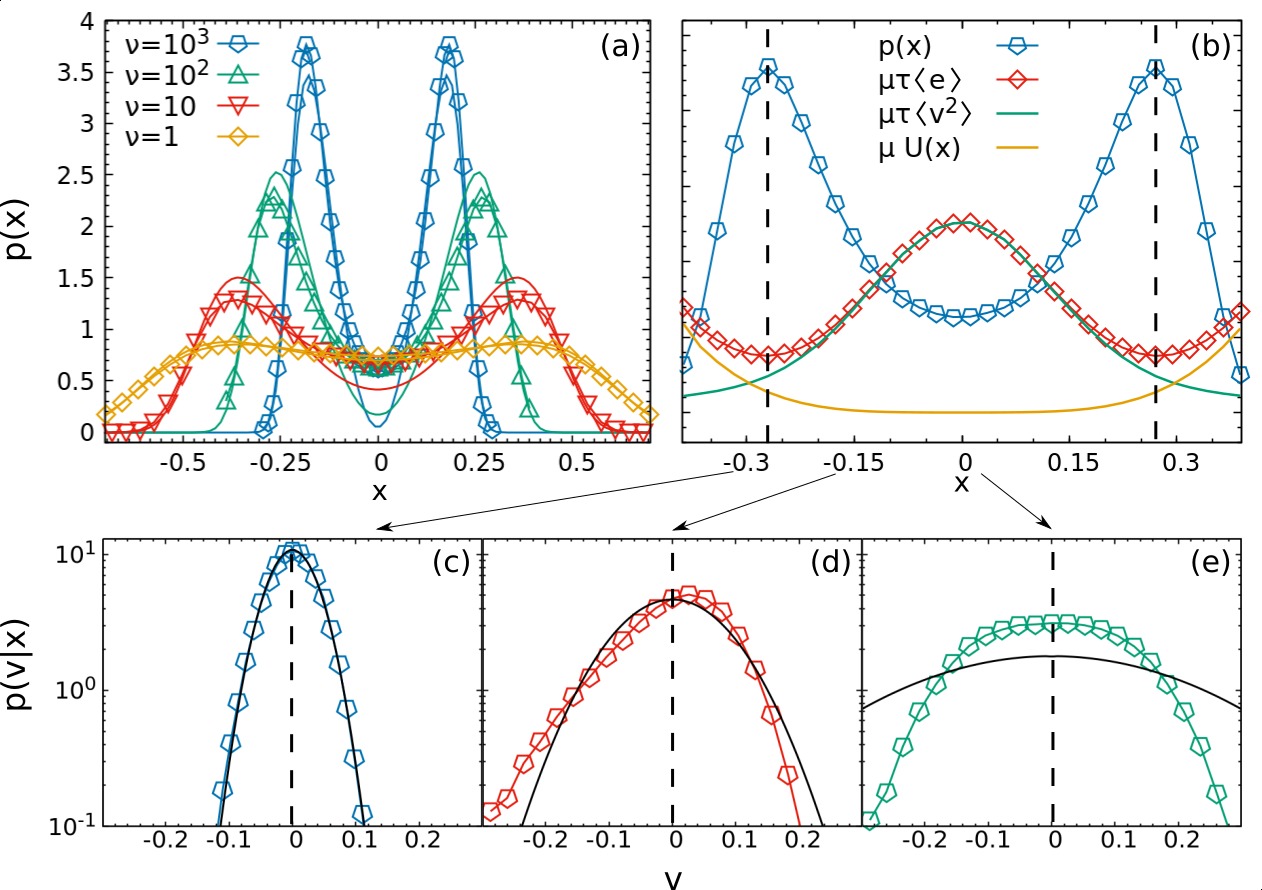}
\caption{Top Panel:  on the left $p(x)$ computed from data (line+dot) and $p_u$ (line), for different values of $\tau$.
On the right, for $\tau=10$: $p(x)$ (blue points), energy $\langle e \rangle_x$ (red points), $\langle v^2\rangle_x$ (green line) and $U(x)$ (orange line). Two vertical black lines are drawn at $x=x_m$, corresponding to the most probable position. Bottom Panel: $p(v|x)$ for three different positions, $x=-0.3, -0.15, 0$, from left to right. The black line is the equilibrium prediction.
Parameters: $D_a=1$, $\tau=10$, $k=10/4$, $n=2$. }\label{fig:EnergyUCNA}
\end{figure}

We also consider a hydrodynamic approach, explained in details in ref.~\cite{marconi2017heat}, which provides a useful tool to improve the understanding of the observed phenomenon. In particular, let us start from the Fokker Planck (FP) Equation associated to Eqs.\eqref{eq:xv_model}, in 1d in the non-interacting case. Multiplying by a polynomial basis in powers of $v$ and integrating out the velocity, we can construct an infinite herarchy of equations, involving the probability density $p(x)$, the first velocity momentum, $\langle v\rangle_x$, the second velocity momentum, $\langle v^2 \rangle_x$, and so on. Here, we have introduced the notation $\langle \cdot \rangle_x=\int dv\cdot p(x,v)/p(x)$, which points out that each observable is an explicit function of the position $x$. The zero-order equation, obtained just by integrating out the velocity in the FP Eq., is equivalent to the mass conservation.
 The first order equation obtained by multiplying FP by $v$ and integrating out the velocity reads:
\begin{equation}
\label{eq:hydro_firstorder}
\frac{\partial}{\partial t}[p(x)\langle v\rangle_x]+\frac{\partial}{\partial x}[p(x)\langle v^2 \rangle_x] =\left[ \frac{F}{\mu} -\frac{\gamma\,\Gamma}{\mu}\langle v\rangle_x\right] p(x) .
\end{equation}
Eq.\eqref{eq:hydro_firstorder} expresses the evolution of the particles momentum, in terms of $\langle v^2\rangle_x$ and $p(x)$. Note that $\langle v^2\rangle_x$ is not constant in space, at variance with ordinary underdamped equilibrium dynamics. Iterating this procedure in the polynomial $v$-basis leads to an infinite hyerarchy of equations, which cannot be solved without employing some closure.
 Since in the stationary state $\langle
v\rangle_x=0$, the minima of $p(x)$ correspond to the minima of the
function $\langle e \rangle_x= \langle v^2\rangle_x+U/\mu$. The slowdown of the particles in regions far from the minima balances the
increasing in the potential energy. These results are well verified
in Fig.~\ref{fig:EnergyUCNA}~{(b)}. Let us notice that the space dependence
of $\langle v^2 \rangle_x$ is determined by the correlation between $x$
and $v$  and connected with the violation of the detailed balance
condition and of the equipartition theorem \cite{fodor2016far,Maggi2017FluctuationRelation}.

\begin{figure}[!t]
\centering
\includegraphics[width=1\linewidth,keepaspectratio]{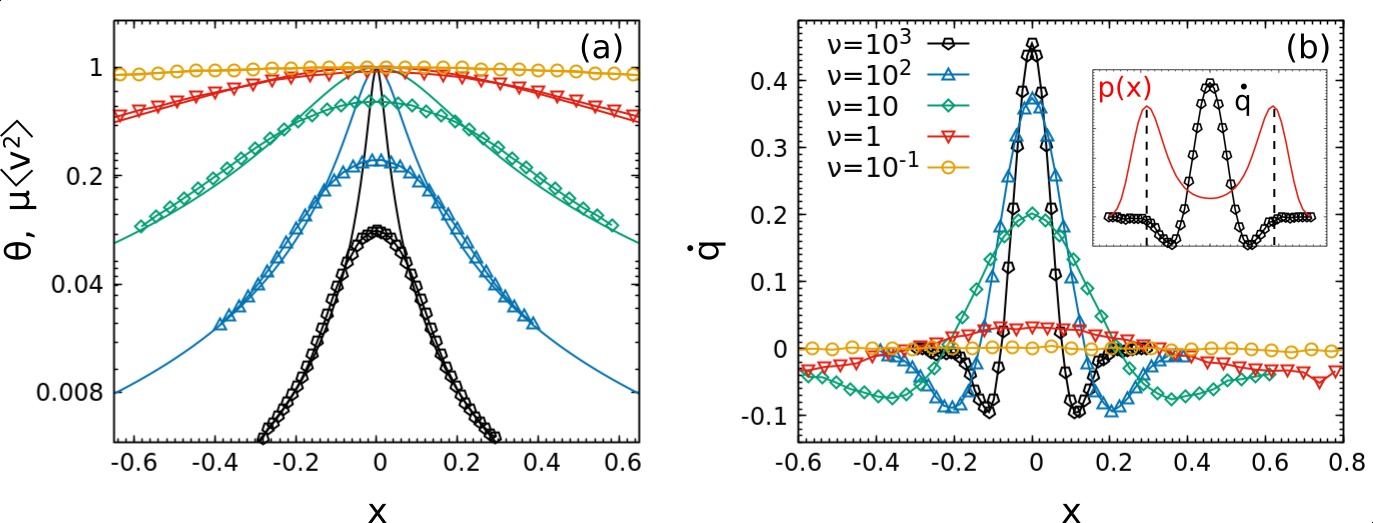}
\caption{ Panel (a): Temperatures $\theta(x)$ (line) and $\mu\langle v^2\rangle_x$ (line+dot) in function of $x$.
Panel (b): $\dot{q}(x)$,  in unit of $D_a/\tau$, for different values of $\tau$. Data are collected through a numerical simulation performed with $N=10^4$ independent particles. Parameters: $k=10/4$, $n=2$.}\label{fig:TemperatureHeat}
\end{figure}

\subsection*{Heat, temperature and local detailed balance}
The last observation suggests the existence of non-trivial
thermodynamics balances in this system. The analysis of
Eqs.~\eqref{eq:xv_model} shows that  additional temperature scales
exist, which are space-dependent. Their definitions are clear for one
particle in one dimension, where Eq.~\eqref{eq:v} without external
potential takes the form of an equilibrium bath at temperature
$\theta(x)=T_a/\Gamma(x)$. In the multidimensional case the symmetric
matrix $\Gamma$ can be diagonalised and one obtains a vector of
temperatures~\cite{Puglisi2017Clausius} (for instance in the radial $2d$ case one has a radial
temperature and a tangential temperature). In~\cite{marconi2017heat,Puglisi2017Clausius} it was shown
that such a temperature satisfies a generalized Clausius relation
coupling entropy production and heat exchanged with the bath.  In
particular, following a stochastic thermodynamics approach \cite{GC1995,LS1999,S2005,SS2005,SS2006},
the entropy production rate of the medium $\dot{S}_m$ can be calculated. Despite the recent dispute about
$\dot{S}_m$, the validity of the result was definitively confirmed in \cite{CommentEntropyProd}.
 Moreover, $\dot{S}_m$ and the heat
rate density, $\dot{q}(x)$, in 1D, are related through the relation~\cite{Puglisi2017Clausius}:
\begin{flalign}
  \dot{S}_m&= \int dx\, p(x)\dot{q}(x)/\theta(x),\\
\dot{q}(x)&=  \frac{D_a\gamma}{\tau\,\theta(x)} \left[\theta(x) - \mu\left< v^2 \right>_x  \right], \;\;\;\;\;\;
\theta(x)= \frac{D_a\gamma}{\Gamma}=D_a\gamma \left(1+\frac{\tau}{\gamma}U''(x)\right)^{-1}.
\end{flalign}
Physically speaking, at $x$ a local flux of heat is transferred from the system to
the active bath if $(\mu \langle v^2 \rangle_x
-\theta(x))$ is positive while the reverse occurs in the negative case.  In the Fig.~\ref{fig:TemperatureHeat}~(a),
we numerically compare the temperature $\theta$ and $\mu\langle v^2
\rangle_x$, showing a clear discrepancy in the central part of the
system which increases with $\tau$. Interestingly, both temperatures
rapidly decrease when moving from the origin to the periphery of the
well, making it clear that the annular region where density is high is
also very cold. In the proximity of highest density, we have $\mu
\langle v^2 \rangle_x \sim \theta(x)$, meaning that in that region the particles reach an effective thermal equilibrium 
with the heat bath so that the 
DB is locally satisfied, although globally it is not. This picture is
confirmed by Fig.~\ref{fig:TemperatureHeat}~(b) where the local
exchange of heat is shown, becoming negligible in the positions
corresponding to the density maxima. Therefore, we can identify two
symmetric space regions (ER), where the system is almost in equilibrium
and others where it is strongly far from it (NER).  In order to
confirm our intuition, we plot the local conditional probability, $p(v|x)=p(x,v)/p(x)$, in the bottom graphs
of Fig.~\ref{fig:EnergyUCNA} (Panels (c)~-~(d)). The Gaussian prediction at temperature $\theta(x)$ in the ER and a
strongly non-Gaussian shape in the NER are confirmed: going towards
the origin, $p(v|x)$ becomes an asymmetric function with a skewed tail
until the symmetrization is again reached in $x=0$, where the non
Gaussianity is still quite clear. Comparing $p(x)$ and $p(v|x)$, we
note that a particle spends most of its time in the ER, where it
accumulates a small amount of heat per unit of time through the
coupling with the fictitious bath. When a fluctuation gives it enough energy, it can
overcome the effective barrier which separates the two effective
symmetric wells, rapidly crossing the NER, and rapidly returning all the heat,
absorbed before, to the bath (indeed numerically $\int dx\,p(x) \,\dot{q}=0$),
in order to come back in the ER.

\section*{Summary and Conclusion}
In conclusion, we have reproduced the recent experimental observation
of the delocalization phenomenon by means of a simple model of
self-propelled particles.  We showed that interactions do
  not suppress the phenomenon (unless close packing is reached) but
  may induce interesting internal structures which, when
  self-propulsion is relevant, can be hardly captured by equilibrium
  modeling and are sensitive to changes of activity time.
  Interestingly, in the delocalized regime, a local detailed balance
  condition is verified in the preferred regions.  Our conjecture is
  that this is the reason why the peaks of the density distribution
  are fairly reproduced by the UCNA approximation in terms of an
  effective double well potential and an equilibrium-like approach
  works \cite{Wittman2017Effective}. Escape times through the
effective double well potential could be interesting and improve
previous studies~\cite{sharma2016,sharma2017} where the authors found
just a polynomial correction to the Kramers-formula~\cite{Gardiner}.




\section*{Contributions}

L.C., U.M.B.M., and A.P. contributed equally to the manuscript.

\section*{Additional information. Competing interests}

The authors declare no competing financial and non-financial interests.







\section{Supplemental Material}

In this Supplemental Material, we shall discuss in more detail with respect to the main text (MT) the phenomenology illustrated concerning the radial pair correlation function, $g(r)$, and  the concept of effective temperatures of an active system~\cite{bechinger2016active,ramaswamy2010mechanics,marchetti2013hydrodynamics}. {Moreover, we present the derivations of the approximations employed in MT. }
{In particular, in Sec.~\ref{sec:DerEq3} we derive Eqs.(3) starting from Eqs.(2) of MT and the Unified Colored Noise Approximation (UCNA), i.e. Eq.(4) of MT.}
In Sec.~\ref{sec:T}, we discuss the concept of effective temperature applied to the AOUP system~\cite{szamel2014self,maggi2015multidimensional,marconi2015velocity,Gompper2018}.
In particular, we discuss whether to characterize the system it is appropriate to define the effective temperature, $T$, through the Gibbs density configurational distribution  $ \propto e^{ -U(r)/T_{eff}}$ or we need alternative definitions, for instance by identifying $T$ with the average kinetic energy of the particles. 
In  Sec.~\ref{sec:gr}, we study in detail the $g(r)$ of the system, 
explaining the computational details and performing an extensive comparison with a Brownian system under the same conditions of density and temperature.

\section{Derivation of Eq.(3) \& UCNA-approximation}
\label{sec:DerEq3}
{
In this Section we review for the sake of completeness the derivation of Eqs.(3) of the main text (see refs.\cite{marconi2015velocity,marconi2015towards}).
Let us start from Eqs.(3) of MT, describing the interacting dynamics of AOUP active particles. Neglecting the thermal noise, these equations read (using Cartesian components and Einstein's summation convention):
\begin{flalign}
\label{eq:AOUPxu_x}
\dot{x}_{i\alpha} &= - \frac{\partial_{i\alpha} \Phi}{\gamma} + u^a_{i\alpha}\\
\label{eq:AOUPxu_u}
\dot{u}^a_{i\alpha}&= -\frac{u^a_{i\alpha}}{\tau} + \frac{\sqrt{2D_a}}{\tau}\eta_{i\alpha}
\end{flalign}
where $\Psi$ is the total potential acting on the system. 
The Latin index identifies the particles  and the Greek index specify the Cartesian component  of each vector. Applying the time-derivative to Eq.\eqref{eq:AOUPxu_x} and defining the coarse-grained velocity:
\begin{flalign}
\label{eq:vdef}
v_{i\alpha}=\dot{x}_{i\alpha} ,
\end{flalign}
we obtain:
\begin{flalign}
\ddot{x}_{i\alpha} &= - \frac{v_{j\beta}}{\gamma}\partial_{j\beta}\partial_{i\alpha} \Phi + \dot{u}^a_{i\alpha}=- \frac{v_{j\beta}}{\gamma}\partial_{j\beta}\partial_{i\alpha} \Phi -\frac{u^a_{i\alpha}}{\tau} + \frac{\sqrt{2D_a}}{\tau}\eta_{i\alpha}
\end{flalign}
where in the last equality we have used Eq.\eqref{eq:AOUPxu_u} to eliminate $\dot{u}^a_{i\alpha}$. Now, using Eq.\eqref{eq:AOUPxu_x} and \eqref{eq:vdef} to eliminate $u^a_{i\alpha}$ we obtain:
\begin{flalign}
\label{eq:AOUPxv_v}
\dot{v}_{i\alpha} = - \frac{v_{j\beta}}{\gamma}\partial_{j\beta}\partial_{i\alpha} \Phi - \frac{\partial_{i\alpha}\Psi}{\gamma\tau} - \frac{v_{i\alpha}}{\tau}+ \frac{\sqrt{2D_a}}{\tau}\eta_{i\alpha} = - \frac{\partial_{i\alpha}\Psi}{\gamma\tau} - \frac{1}{\tau} \left[ \delta_{ij}\delta_{\alpha\beta}  + \frac{\tau}{\gamma}\partial_{j\beta}\partial_{i\alpha} \Phi \right] v_{j\beta} + \frac{\sqrt{2D_a}}{\tau}\eta_{i\alpha}
\end{flalign}
which is Eqs.(3) of MT, being $\Gamma_{ij\alpha\beta} = \delta_{ij}\delta_{\alpha\beta}  + \frac{\tau}{\gamma}\partial_{j\beta}\partial_{i\alpha} \Phi$.

We can derive the Unified colored noise approximation (UCNA)  by taking the over-damped limit, $\dot{v}_{i\alpha}\approx 0$, in Eq.\eqref{eq:AOUPxv_v}. This procedure leads to a relation between $v$ and $x$:
\begin{equation}
\label{eq:UCNA_dynamics}
\dot{x}_{i\alpha} = v_{i\alpha} = \Gamma^{-1}_{ij\alpha\beta}\left[ -  \frac{\partial_{j\beta}\Psi}{\gamma} +  \sqrt{2D_a}\eta_{j\beta}\right]
\end{equation}
ruling the UCNA dynamics. Eq.\eqref{eq:UCNA_dynamics} involves a multiplicative noise and does not satisfy the fluctuation-dissipation theorem. The associated Fokker-Planck equation associated to Eq.\eqref{eq:UCNA_dynamics} - using the Stratonovich integration - , for the density $p(\{x\},t)$, reads:
\begin{equation}
\partial_t p = - \partial_{i\alpha}J_{i\alpha}, \qquad J_{i\alpha}= -\Gamma^{-1}_{i\alpha k\beta} \left[\partial_{k\beta}\Psi \,\,p+ D_a \partial_{k\beta}\left( \Gamma^{-1}_{i\alpha j\beta}\,\,p \right)    \right]
\label{ucnaeq}
\end{equation}
Looking for a stationary solution of Eq.~\eqref{ucnaeq} under the condition of vanishing current  we find the following equation:
\begin{equation}
-p\,\,\frac{\partial_{k\alpha}\Psi}{D_a \gamma}  - p\,\,\partial_{j\beta} \Gamma^{-1}_{j\beta i\alpha} = \Gamma_{j\beta i\alpha}[\partial_{j\beta} p] .
\end{equation}
After some algebra we obtain:
\begin{equation}
\label{eq:zeroCurrent_UCNA}
-p\,\, \frac{\tau}{D_a\gamma} \Gamma_{i\alpha j\beta}\partial_{j\beta} \Psi + p \,\,\Gamma^{-1}_{j\beta k \delta}\partial_{j\beta}\Gamma_{k\delta i \alpha} = \partial_{i\alpha} p
\end{equation}
Moreover, by using the identity, $\partial_{j\beta}\Gamma_{i\alpha k\gamma}= \partial_{k\gamma}\Gamma_{i\alpha j\beta}$ - since $\Gamma$ involves only the second derivatives of $\Psi$ -  and the Jacobi's formula:
\begin{equation}
\frac{1}{\det{\Gamma}} \partial_y \det{\Gamma} = Tr\left( \Gamma^{-1}\partial_y \Gamma  \right)
\end{equation}
we obtain:
\begin{equation}
\Gamma^{-1}_{i\alpha j \beta} \partial_{ i \alpha} \Gamma_{j\beta k\delta} = \frac{1}{\det{\Gamma}} \partial_{k\delta} \det{\Gamma}
\end{equation}
Using this result in Eq.\eqref{eq:zeroCurrent_UCNA}:
\begin{equation}
-D_a\gamma \left( \partial_{i\alpha} p   - p\,\,\partial_{i\alpha} \log{\det{\Gamma}}  \right) = p\,\, \Gamma_{i\alpha j\beta}\partial_{k\beta} \Psi
\end{equation}
Solving this set of partial first-order differential Eqs. we find the UCNA probability distribution:
\begin{equation}
\begin{aligned}
p &\propto \exp{\left(-\frac{\mathcal{H}}{D_a \gamma}\right)},\\
\mathcal{H} &= \Psi + \frac{\tau}{2\gamma} \sum_{k\beta}\left(\partial_{k\beta} \Psi\right)^2 - D_a\gamma \log{\det{\Gamma}} .
\end{aligned}
\end{equation}
Considering just the one-dimensional case in the non-interacting case, we find the pdf $p(x)$ shown in Fig.2 of MT

The possibility of neglecting $\dot{v}=0$ in Eq.\eqref{eq:AOUPxv_v},  i.e. taking the overdamped limit, is equivalent to assume the Gaussianity of the conditional probability, $p(v|x)$, with a kinetic "effective temperature" which satisfies the Einstein relation. The form of such a $p(v|x)$  shows that  different components of the velocity are not independent. In fact, the approximate probability distribution reads:
\begin{equation}
\label{eq:approxprob}
p(\{x\},\{v\}) = p(\{v\}|\{x\})p_u(\{x\}) \propto e^{-v_{i\alpha} \Gamma_{ij\alpha\beta}v_{j\beta}} p_u .
\end{equation}
We point out that Eq.\eqref{eq:approxprob} is not the solution of the FP-equation associated to Eq.\eqref{eq:AOUPxv_v}, but can be just considered as a useful approximation. Despite its apparent simplicity, it involves many-body interactions, which cannot be easily evaluated and for this reason, up to now, the UCNA was not particularly practical in understanding collective phenomena.}

\section{A kinetic temperature for the active system}
\label{sec:T}
In the presence of an external potential, it is not clear which should be the temperature of an assembly of active particles~\cite{Puglisi2017Temperatures}.
Recently, some approximations were developed with the aim of describing by an effective potential the particles interactions~\cite{h1989colored,hanggi1995colored,marconi2015towards,Wittman2017Effective}. These approximations seem to work in spite of the fact that these systems are clearly far from equilibrium~\cite{FodorMarchetti2018,fodor2016far}.

 As discussed in MT, the potential-free system displays two temperatures: $T_b=\gamma D_t$ determined by the solvent, and $T_a=\gamma D_a$, (the so called active temperature) related to the self-propulsion force, $\gamma u$. Since we fix the variance of $u$, i.e. the ratio $D_a/\tau$, the possibility of neglecting $T_b$ with respect to $T_a$ depends on the value of $\tau$. { On one hand, for $\tau$ small enough $T_a\ll T_b$, one encounters a non-interesting regime where the activity plays a negligible role and the system behaves as if it were subject to Brownian dynamics, at temperature $T_b$. On the other hand, the more interesting regime studied in the main text
occurs when $\tau$ is large enough,  i.e. $T_a \gg T_b$, so that we can effectively neglect the solvent temperature.}
The presence of a non uniform external force leads to a new effect: it determines a non-trivial correlation between the position, $x$, of the particle and its self-propulsion, $\gamma u$, which eventually leads to the violation of the equipartition theorem \cite{fodor2016far,Maggi2017FluctuationRelation}, breaking the DB \cite{marconi2017heat}.
{ In this case, the identification of $T_a$ with an effective temperature~\cite{Puglisi2017Temperatures,Cugliandolo2011EffectiveT} is not trivial and depending on the choice of the parameters
in general not true. In MT, we exploit the importance of $\nu$ - i.e. the ratio between the typical time 
$\tau$ associated with the active force and the one associated with the potential, $\gamma/ U''(l)$}. The dimensionless
parameter $\nu$ is recognized as the relevant parameter determining if the system is close to a global equilibrium. In particular, if $\nu\ll1$ but $T_a \gg T_b$ we can perform the overdamped limit of the Eq.1(a) of MT, approximating $u$ as a Brownian process. This operation provides a simplified overdamped dynamics for the particle position, meaning that the system reaches the equilibrium, evolving with an effective Brownian dynamics with diffusion coefficient, $D_a$:
\begin{equation}
\dot{x}=-\frac{U'(x)}{\gamma} + \sqrt{2D_a}\eta \qquad\longrightarrow \qquad p(x) \propto \exp{\left[-\frac{U(x)}{\gamma D_a} \right]}.
\end{equation}
In this regime, $T_a=\gamma D_a$ has, trivially, the role of the effective temperature of the system. For $\nu \sim O(1)$, this is no longer true, since { the system is not in the overdamped regime}. Moreover, we can directly check this claim by evaluating a simple solvable case: the harmonic potential in one dimension. Indeed, by setting $U(x)=k x^2/2$, Eqs.(2) 
of MT can be solved~\cite{szamel2014self,Gompper2018}, providing an analytical expression for the steady state probability $p(x,u)$ for all values of $\nu$:
\begin{equation}
p(x, u) \propto \exp{\left[-\frac{k}{\gamma}\frac{\Gamma}{D_a}\frac{x^2}{2} \right]}\exp{\left[- \frac{\tau}{D_a}\frac{\Gamma}{2} \left( u - \frac{k}{\gamma}x \right)^2   \right]}, \qquad \Gamma=1+\nu
\end{equation}   
As we can see, $T_a$ does not coincide with the effective temperature of the system when $\nu$ is not negligible, a result which is in general true for a generic potential, except for some special cases~\cite{szamel2014self}. 

{ The  UCNA equilibrium-like approach, employed} in MT and reviewed in Sec.\ref{sec:DerEq3}, provides a prediction for the equilibrium temperature $\theta(x)$ in the non-interacting one dimensional system: $\theta(x)=\gamma D_a / \Gamma(x)$, being $\Gamma(x)=1+U''(x)\tau/\gamma$. {In particular, we find numerically that UCNA does not hold }globally  in space but only in the so-called equilibrium regions (ER), which correspond to the regions where the particles spend most of their life, as shown in MT. In the ER the stationary probability distribution, $p(x,v)$, is a Gaussian with respect to $v$:
\begin{equation}
p(x,v)\propto e^{-H_u/D_a\gamma} e^{- \mu v^2/2\theta(x)}, \qquad H_u= U(x) + \frac{\tau}{2\gamma}\left[\frac{\partial}{\partial x} U(x)\right]^2 - \gamma D_a\log{\left[\Gamma(x)\right]},
\end{equation} 
{adapting Eq.\eqref{eq:approxprob} to the one-dimensional non-interacting case. } { Therefore,}  $\theta(x)$ can be interpreted as a space dependent kinetic temperature.
For small activity, $\theta$  is almost equivalent to $T_a$, but this is no longer true at large activity {since the space-dependence plays an important role.}

\section{Spatial structure and thermal equivalents of the active system}\label{sec:gr}

In Fig.1 of the main text, we have reported the important changes of $g(r)$ 
when $R/l$ and $\nu$ are varied. 
Here, we discuss the possibility of interpreting these changes in terms of some effective temperature.
Let us see what happens to $\theta(x)$ which, according to Section 1, can be interpreted as an effective local kinetic temperature.

As discussed in Section 1, $\theta$ scales as $ D_a\gamma/\tau$ when $\nu\gg1$, meaning that the increasing of $\nu$, leaves unchanged this effective temperature of the system since the ratio $D_a/\tau$ is fixed.
In the bottom panel of Fig.(1) of the main text, we display $g(r)$ for different interaction lengths, $R/l$, and for two values of $\nu=10^2, 10^4$. In all cases, a freezing phenomenon seems to occur with the increase of $\nu$, since the peaks of $g(r)$ become more pronounced.
We point out that, on one hand, these measures were performed by monitoring the effective density of the system in the more crowded regions: $\rho$ remains nearly constant in such a way that its variation cannot be considered the cause of the structural changes appearing in $g(r)$.  Also  $\theta$ roughly does not change, meaning that such structural changes are not driven by a variation of $\theta$.

We try an alternative approach and look whether an equivalent Brownian system exists - at the same density and  appropriate temperature - displaying the same $g(r)$. To answer these questions boils down to establish whether there exists or not  a mapping between the active system and a fictitious over damped passive system.

For an equilibrium system of passive interacting Brownian particles ($D_a=0$), with diffusion coefficient $D_t$, the Einstein relation holds and we can identify $\gamma D_t $ as the temperature of the system, $T_b$. At fixed area fraction, a variation of $T_b$ produces a change in the structure of the system, which can be analyzed by the pair correlation function, $g(r)$~\cite{lowen1994review}. For $T_b$ large enough the $g(r)$ is flat, meaning that there are not preferential distances among particles, a situation which can be roughly identified as a gas-like phase. Particles move around the accessible volume and the interactions are rare and binary-like. The decreasing of $T_b$, produces some peaks in the $g(r)$, before approaching to one. These peaks establish the typical distances among particles, a regime identified as liquid-like, since particles move around the available volume and particles positions are strongly correlated.
A further decreasing of $T_b$ leads to a freezing pattern, where particles just fluctuate around their fixed equilibrium positions, which are hexagonally distributed in two dimensions.
The $g(r)$ peaks become higher and thin, approaching to $\delta$-Dirac function (ideally at $T_b=0$), a phase which has strong analogies with a solid. 
The same qualitative picture, gas $\rightarrow$ liquid $\rightarrow$ solid, is, roughly, produced by the growth of the packing fraction of the system, $\phi\approx\rho_0 R^2$ (in two dimensions), being $\rho_0$ the numerical density and $R$ the interaction length of the repulsive pairwise potential. 
Let's remark that the identification of the microscopic structures with the macroscopic phases (eventually with phase transitions) makes sense just if we consider the infinite volume limit. If  we apply an external potential, particles can explore just an effective volume, depending both on the inter-particle interactions and on the potential itself. 
Therefore, fixing the number of particles to $\sim 10^{2} - 10^{4}$ (typical numbers of a simulation), means to study a system with few degrees of freedom, whose importance, nowadays, is well known. With these motivations, studying the internal structure of such a system, for instance through the $g(r)$, makes sense and could be useful in order to understand the role of the interactions.

In Fig.~1 of the SM we compare the $g(r)$ of active systems with that of many possible passive systems
having the same interactions and density (for details see Sec.~\ref{sec:computation}),  varying the diffusion coefficient.
From this analysis it emerges that in the gas-like regime (Fig.~1 panel (d)), when $R$ is very small, $g(r)$ displays a first peak at $r\approx R$, which does not have a Brownian counterpart. Indeed, the passive $g(r)$ profiles are flat also for very low temperatures, which is not a surprise since the interactions are rare. We interpret the active peak as a consequence of the slow-down of the particles in the presence of a convex interaction, as $\sim 1/|\mathbf{x}|^b$ with $b>0$, which increases the probability that two particles are close to each other. This is the leading mechanism on which the MIPS phenomenon is based.

\begin{figure}[!t]
\centering
\includegraphics[width=1\linewidth,keepaspectratio]{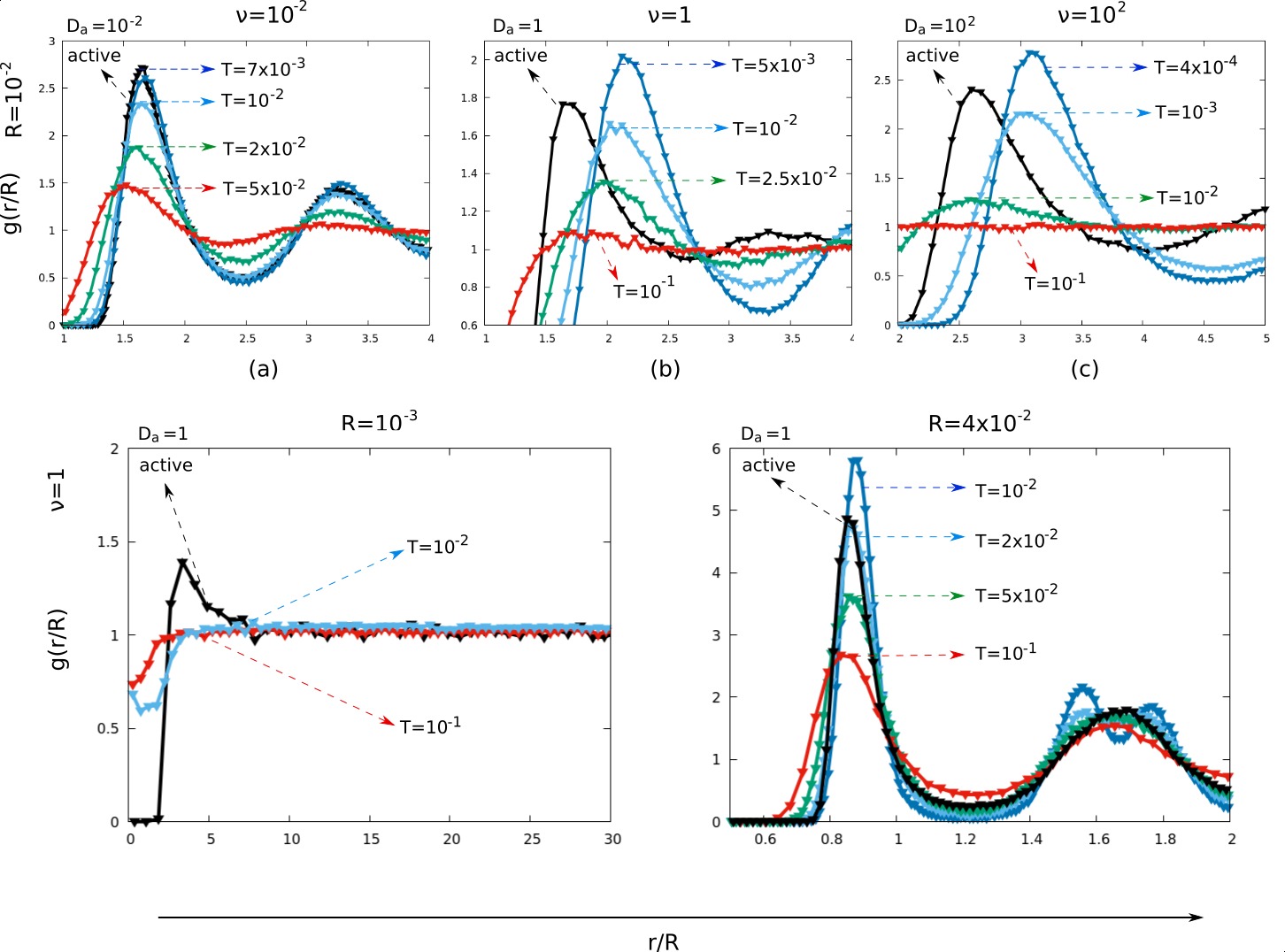}
\caption{$g(r)$ for an active system (black line) compared with the equivalent Brownian systems at different temperatures (colored line as shown in each graph). Top Panel: $R=10^{-2}$, graph(a),(b) and (c), respectively, at $\nu=10^{-2},1,10^{2}$. Bottom panels: $\nu=1$. Graph(d) at $R=10^{-3}$ and graph(e) at $R=4\cdot10^{-2}$. Other parameters: $k=10^2/4$, $D_a/\tau=10^2$, $b=4$. }\label{fig:gr_Brownian}
\end{figure}

In the top panel of Fig.~\ref{fig:gr_Brownian}, we perform the same analysis with an interaction $R$ larger than the previous case, such that a liquid-like structure is  produced as the pronounced peaks reveal. 
For $\nu \geq 1$,
 as shown in panels (b) and (c), again it is not possible to determine a value of the diffusion coefficient in such a way that we can reproduce the shape of $g(r)$ in the active case: indeed, the active peak is always shifted towards smaller values of $r$. Such an effect clearly disappears when $\nu \ll 1$ since in the last case an active system is equivalent to a passive overdamped one with  effective temperature $T_a$ (Fig~\ref{fig:gr_Brownian} panel (a)).

Finally, a further increasing of $R$, leads to a completely different scenario also for $\nu \sim O(1)$. Although the effective temperature of the system does not trivially scale  with $\nu$, the comparison with a Brownian simulation shows that we can find a numerical temperature value, $T_r$, through which we can reproduce the active $g(r)$ shape, as illustrated in Fig.~\ref{fig:gr_Brownian} panel (e).

We may conclude that when the packing fraction is large enough the microscopic structure, represented by $g(r)$, is the same as the one of an equivalent Brownian system with an effective temperature, $T_r$. Nevertheless, there is no way to reproduce  the active $g(r)$ in the gas-like and liquid-like regime induced by the activity, with an equivalent Brownian simulation under the same condition: these structural changes are entirely due to the activity and are genuine non-equilibrium effects.

\subsection{Details about the $g(r)$ computations}\label{sec:computation}
In the active case, we have computed numerically the $g(r)$ function by using the following procedure:
\begin{enumerate}
\item We chose a square inside the dense region of a configuration: a disk or a circular crown (depending if the radial delocalization occurs or not). This square is chosen not too big, in such a way we may neglect the boundary of such a region.
\item We compute numerically the $g(r)$ in this region, using the standard formula~\cite{lowen1994review}:
 $$g(r)=\frac{V}{N}\left\langle\frac{1}{N} \sum_{i\neq j}\delta[\bm{r}- (\bm{r}_i-\bm{r}_j)] \right\rangle$$,
where $\bm{r}_i$ is the position of a target particle and $\langle\cdot \rangle$ denotes both the average over all the particles inside the region and a time average. The normalization is estimated by numerically computing the number of particles inside the square for each configuration.
\item Consistency check: in order to check the result, we perform the same analysis for different (but dense) regions, verifying that there are no big changes.
\end{enumerate}

In the following, we describe the protocol adopted in order to compare an AOUP system with a Brownian one. In particular, we are interested in performing a Brownian simulation under the same conditions of an active system, i.e. same packing fraction and number of particles:
\begin{enumerate}
\item We compute the $g(r)$ in the active case, with the procedure described above, by selecting a square space region, of area $A_r$, and computing the average number of particles, $N_r$, in that region. 
\item We compute the $g(r)$, for the following system: $N_r$ overdamped Brownian interacting particles in a square region of area, $A_r$, under the action of the confining potential. In this way, the numerical density of the equivalent Brownian system is the same as the AOUP. Considering the same interactions we have a Brownian system with the same packing fraction. 
\item We compute the $g(r)$ of the Brownian system for different values of the diffusion coefficient, checking if there is some temperature value which reproduces the active $g(r)$.
\end{enumerate}

\end{document}